# A Compact Virtual-Source Model for Carbon Nanotube Field-Effect Transistors in the Sub-10-nm Regime—Part II: Extrinsic Elements, Performance Assessment, and Design Optimization

Chi-Shuen Lee, Eric Pop, *Senior Member, IEEE*, Aaron D. Franklin, *Senior Member*, *IEEE*, Wilfried Haensch, *Fellow, IEEE*, and H.-S. Philip Wong, *Fellow, IEEE*

*Abstract*— We present a data-calibrated compact model of carbon nanotube (CNT) field-effect transistors (CNFETs) including contact resistance, direct source-to-drain and band-to-band tunneling currents. The model captures the effects of dimensional scaling and performance degradations due to parasitic effects and is used to study the trade-offs between the drive current and leakage current of CNFETs according to the selection of CNT diameter, CNT density, contact length, and gate length for a target contacted gate pitch. We describe a co-optimization study of CNFET device parameters near the limits of scaling with physical insight, and project the CNFET performance at the 5-nm technology node with an estimated contacted gate pitch of 31 nm. Based on the analysis including parasitic resistance, capacitance, and tunneling leakage current, a CNT density of 180 CNTs/μm will enable CNFET technology to meet the ITRS target of drive current (1.33 mA/μm), which is within reach of modern experimental capabilities.

## I. Introduction

Semiconducting single-walled carbon nanotube (CNT) field-effect transistors (CNFETs) have shown promise for extending CMOS technology scaling into the sub-10-nm technology nodes [1-3] owing to CNTs' near-ballistic carrier transport [4-5] and ultra-thin body (1-2 nm), which provides superior electrostatic control over the channel and enables further scaling of the gate length ($L_g$) below 10 nm [3,6]. While CNFETs have superior intrinsic electronic properties, they suffer from imperfections, such as the difficulty of acquiring extremely high-purity semiconducting CNTs [7], hysteresis of the current-voltage (I-V) characteristics [8], and variations of material and devices [9]. Techniques to overcome these imperfections at the system level have been reported in [10] at modest cost of area and energy consumption.

In this paper, we focus on two specific issues: parasitic metal-CNT contact resistance ($R_c$) and direct source-to-drain tunneling (SDT) current ($I_{SDT}$)[1]. Obtaining low $R_c$ between metals and low-dimensional materials has been recognized as one of the most challenging yet critical requirements for high performance transistors [11-12]. Furthermore, as $L_g$ scales below 10 nm, $I_{SDT}$ may become significant and cause high leakage power [2,13-14]. While previous works employed rigorous yet computationally intensive modeling methods to study these issues [2,15], here we develop analytical models for $R_c$ and $I_{SDT}$ in CNFETs and study their impacts on the device performance. This paper is organized as follows: models for $R_c$ and $I_{SDT}$ calibrated to experiments and numerical simulations are described in Sections II and III, respectively. These extrinsic elements are then integrated with the intrinsic model developed in [16] based on the virtual-source (VS) approach to arrive at a complete VS-CNFET model; in Section IV, CNFET performance is evaluated at the 5-nm technology node corresponding to a contacted gate pitch $L_{pitch}$ = 31 nm and metal-1 pitch $L_{M1}$ = 25.2 nm. By comparing the drive current against the 2013 International Technology Roadmap for Semiconductors (ITRS) target [17], requirements of CNT density for CNFETs is presented as a guide for technology development; in Section V, we discuss the assumptions of the model and analysis as well as suggestions for future experimental works. The models presented in this paer are calibrated to the data from experiments and numerical simulations based on non-equilibrium Green's function (NEGF) quantum transport. Therefore, this work aims to provide realistic insight into the potentials and challenges of the CNFET technology. Due to the limited space, the complete derivation of all the equations is detailed in [31]; here we only discuss the physics and key results.

Manuscript received Mar. xx, 2015. This work was supported in part through the NCN-NEEDS program, which is funded by the National Science Foundation, contract 1227020-EEC, and by the Semiconductor Research Corporation, and through Systems on Nanoscale Information fabriCs (SONIC), one of the six SRC STARnet Centers, sponsored by MARCO and DARPA, as well as the member companies of the Initiative for Nanoscale Materials and Processes (INMP) at Stanford University.

C. -S. Lee, E. Pop, and H.-S. P. Wong are with the Department of Electrical Engineering, Stanford University, Stanford, CA 94305 USA (e-mail: chishuen@stanford.edu; hspwong@stanford.edu).
A. D. Franklin is with the Department of Electrical and Computer Engineering, Duke University, Durham, NC 27708 (e-mail: aaron.franklin@duke.edu).
W. Haensch is with IBM T. J. Watson Research Center, Yorktown Heights, NY 10598 USA (e-mail: whaensch@us.ibm.com).

---

[1] These two challenges are not unique to CNFETs, but are also challenges of all scaled FETs. The simplicity of the CNT band structure makes this a model system for gaining insight into these challenges for other materials as well.

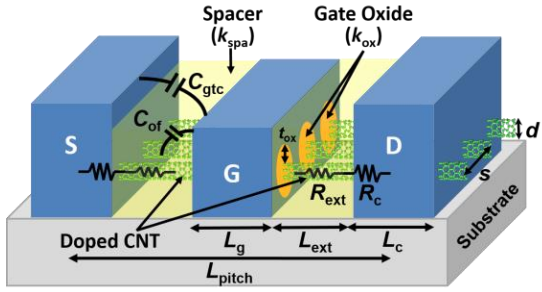

Fig. 1. A representative gate-all-around CNFET structure used in the VS-CNFET model with the critical dimensions, parasitic resistances and capacitances labeled.

## II. PARASITIC RESISTANCE

The CNFET parasitic resistance considered in this work consists of two components: the parasitic metal-CNT contact resistance ($R_c$) and the resistance in the source/drain (S/D) extensions ($R_{ext}$), as illustrated in Fig. 1. In general, metal-CNT $R_c$ is determined by three factors: Schottky barrier height ($\phi_b$), interface quality (i.e. metal-CNT adhesion), and physical contact length ($L_c$). In [18], Fermi-level pinning is predicted to be insignificant in metal-CNT contacts and thus $\phi_b$ is proportional to the CNT band gap ($E_g$) [19]:

$$E_g = \frac{2E_p a_{cc}}{d} \quad (1)$$

where $E_p$ = 3 eV is the tight-binding parameter, $a_{cc}$ = 0.142 nm is the carbon-carbon distance in CNTs, and $d$ is the CNT diameter. Corrections to (1) could be made due to band gap renormalization as discussed in [16], but they do not alter the core of the model presented here. Chen *et al.* experimentally demonstrated an exponential increase in $R_c$ with $1/d$ [20], attributed to the increase in $\phi_b$; other authors showed that lower $R_c$ can be achieved with Pd rather than Au contacts, despite their similar work functions [4,21]. This advantage is attributed to better wettability at the Pd-CNT interface, the importance of which was also clarified by a recent study with several contact metals [22]. In the models presented here we include the dependence of $R_c$ on $d$, but not that of the interface wettability or adhesion (which could also be influenced by polymer residue from fabrication); the dependence of $R_c$ on $L_c$ was experimentally studied in [22-23] and can be phenomenologically modeled by the transmission line model [25]:

$$2R_c = R_Q \sqrt{1 + \frac{4}{\lambda_c g_c R_Q}} \coth\left(\frac{L_c}{L_T}\right) - R_Q \quad (2a)$$

$$L_T = \left[\frac{g_c R_Q}{\lambda_c} + \left(\frac{g_c R_Q}{2}\right)^2\right]^{-1/2} \quad (2b)$$

where $L_T$ is the current transfer length, $R_Q = h/(4q^2) \approx 6.5$ k$\Omega$ is the quantum resistance of the CNT (lowest band, doubly degenerate with two spins), $q$ is the elementary charge, $h$ is Planck's constant, $\lambda_c$ is the charge carrier mean-free-path (MFP) in the CNT under the metal contact, and $g_c$ is the coupling conductance between the CNT and the metal contact. Note that in (2a), $R_Q$ is subtracted on the right-hand side because $R_Q$ is considered the intrinsic property associated with the interfaces between the 1D CNT channel with the metal S/D contacts [24]. As a result, $R_c$ is a parasitic component. In [25], $\lambda_c$ and $g_c$ are constant empirical parameters; whereas in this paper, $g_c$ is related to $\phi_b$ so as to account for the experimental observation of the increase in $R_c$ as $d$ decreases [20] by:

$$g_c = g_{co} \exp(-\phi_b / E_{00}) \quad (3a)$$

$$\phi_b = E_g / 2 - (\phi_m - \phi_s) \quad (3b)$$

where $\phi_m$ and $\phi_s$ are work functions of the contact metal and the CNT, respectively, and $g_{co}$ and $E_{00}$ are empirical parameters. In analogy to the calculation of transmission coefficient through a metal-to-bulk-semiconductor Schottky contact [26], the $E_{00}$ in (3a) characterizes the width of the energy barrier at metal-to-bulk-semiconductor interface: the smaller the $E_{00}$, the wider the barrier, and the more sensitive the $g_c$ to the $\phi_b$. Note that (3b) is for p-type contacts. For n-type contacts, the (3b) should be modified to $\phi_b = E_g/2 + (\phi_m - \phi_s)$.

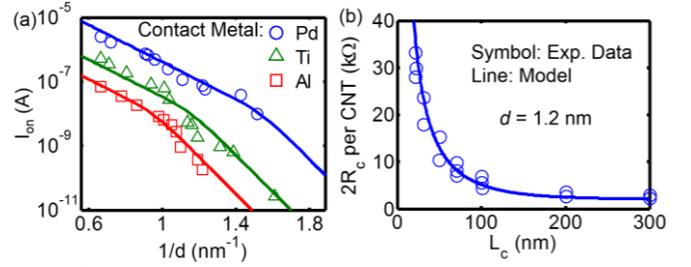

Fig. 2. Parameter extraction for the metal-CNT contact resistance model: (a) $I_{on}$ vs $1/d$ from [20] to extract $E_{00}$ in (3a). (b) $R_c$ vs $L_c$ from [23] to extract $\lambda_c$ and $g_c$ in Eq. (2).

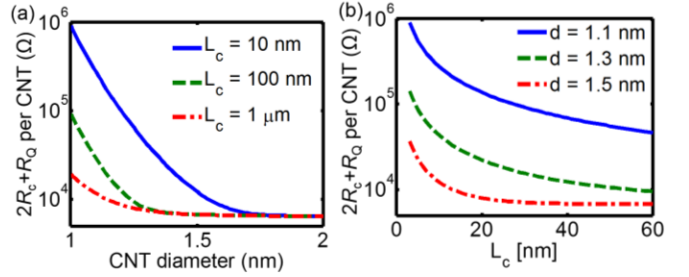

Fig. 3. Contact resistance vs. (a) CNT diameter for different contact lengths and (b) contact lengths for different CNT diameters.

There are three empirical parameters to be determined in (2) and (3): $\lambda_c$, $g_{co}$, and $E_{00}$. The extraction of these three parameters goes as follows: (i) the $R_c$ calculated by (2) and (3) is included into the intrinsic current model described in [16] to generate the on-state current ($I_{on}$) compared against the data from [20] in Fig. 2a. From the slope of $I_{on}$ vs. $1/d$, $E_{00}$ = 32 meV is extracted; (ii) Eq. (2) is fitted to the $R_c$ vs. $L_c$ data from [23] in Fig. 2b, where $\lambda_c$ = 380 nm and $g_c$ = 2 μS/nm are extracted (same as the result in [25]) for $d$ = 1.2 nm with Pd as the contact metal; (iii) substituting $\phi_m$ = 5.1 eV for Pd, $\phi_s$ = 4.7 eV for intrinsic CNTs, $E_g$ = 0.71 eV for $d$ = 1.2 nm, and $g_c$ = 2 μS/nm into Eq. (3a) and (3b), $g_{co}$ = 0.49 μS/nm is obtained. In Fig. 2a, we observe that the $I_{on}$ drops even faster as $1/d$ increases beyond a certain point (for the Al contact as example, the $I_{on}$ decreases more rapidly as $1/d > 1$ nm$^{-1}$). This "accelerated downturn" can be explained as follows: when $1/d$ is small and $g_c$ is large, $L_T \ll L_c$ in (2) and $\coth(L_c/L_T) \approx 1$. Therefore, $R_c$ increases with $(1/g_c)^{1/2} \propto \exp[1/(2d)]$; as $1/d$ increases and $g_c$ becomes small, $L_T \gg L_c$ and $\coth(L_c/L_T) \approx L_T/L_c$, and $R_c$ increases with $1/g_c \propto \exp(1/d)$. This accelerated downturn is observed in both the experimental data and the model (2) and (3), which strengthens the validity of the $R_c$ model. As shown in Fig. 3, in the region where $L_c$ and/or $d$ are small, $R_c$ increases drastically, which severely degrades the drive current and can cause large variation in the presence of variations in $L_c$ and $d$. The impact of $L_c$ and $d$ on the CNFET performance is discussed in Section IV.

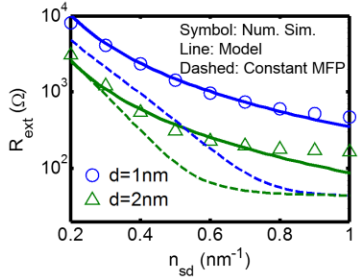

Fig. 4. Comparison of the extension resistances vs. the doping density. The symbols are calculated by (4) numerically and the lines represent the analytical approximation of (5). The dashed lines are generated by assuming $\lambda_i$ in (4) is constant.

The other component $R_{ext}$ is derived from the one-dimensional (1D) Landauer formula [24]:

$$R_{ext} = 1/G - R_Q \quad (4a)$$

$$G = \frac{4q^2}{h} \int_{E_c}^{\infty} \frac{\lambda_i(E)}{L_{ext} + \lambda_i(E)} \left[ -\frac{\partial f(E, E_F)}{\partial E} \right] dE \quad (4b)$$

$$n_{sd} = \int_{E_c}^{\infty} g(E) f(E, E_F) dE \quad (4c)$$

where $G$ is the CNT conductance at low fields, $L_{ext}$ is the length of the S/D extensions (see Fig. 1), $E_c$ is the conduction band edge, $E_F$ is the Fermi level, $E$ is the energy of free electrons referenced to $E_c$, $f$ is the Fermi-Dirac distribution function, $g(E)$ is the CNT density of states (DOS), $n_{sd}$ is the doping density in the S/D extensions, and $\lambda_i$ is the carrier MFP in CNTs representing the aggregate effect of optical phonon and acoustic phonon scattering as introduced in [27]. $R_Q$ is subtracted from $1/G$ in (4a) because $G$ is the total conductance including the contact resistance, which has already been considered in the $R_c$ model. Because $\lambda_i$ has a complex expression [27], Eq. (4b) cannot be integrated analytically. Therefore, an empirical expression of $R_{ext}$ is employed here:

$$R_{ext} = R_{ext0} \frac{L_{ext}}{d^{\alpha_d} n_{sd}^{\alpha_n}} \quad (5)$$

where $R_{ext0}$, $\alpha_d$, and $\alpha_n$ are the empirical fitting parameters. The form of (5) is inspired by the observations that: (i) for heavily doped CNTs, the carrier transport becomes more diffusive and thus $R_{ext} \propto L_{ext}/n_{sd}$ in a manner analogous to the Drude model; (ii) $\lambda_i$ is proportional to $d$ according to [27]. Eq. (5) is then fitted to the numerical results given by (4) as shown in Fig. 4, where $R_{ext0} = 35\ \Omega$, $\alpha_d = 2$, and $\alpha_n = 2.1$ are extracted. Eq. (5) agrees well with (4) at low $n_{sd}$ region but underestimates $R_{ext}$ at high $n_{sd}$ region. However, when $n_{sd}$ is large, $R_{ext} \ll R_c$ so the discrepancy is negligible. The dashed lines in Fig. 4 represent the results when $\lambda_i$ is a constant instead of being dependent on energy and CNT diameter. In such a case, $R_{ext}$ exhibits less sensitivity to $d$ and higher sensitivity to $n_{sd}$.

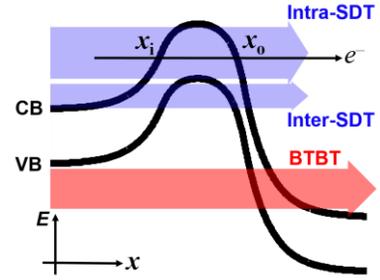

Fig. 5. Illustration of the direct source-to-drain tunneling and the band-to-band tunneling mechanisms. $x_i$ and $x_o$ are the positions where the electrons "tunnel" in and out the energy barrier.

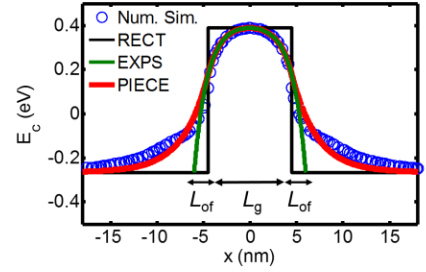

Fig. 6. Conduction band profile calculated by the numerical simulation (circles) [30] and the three analytical models—RECT: a rectangular $E_c$ profile; EXPS: two connected exponential functions given by (9); PIECE: a piecewise function given by (10).

## III. TUNNELING LEAKAGE CURRENT

According to the 2013 ITRS projections [17], the $L_g$ of MOSFETs should eventually scale below 10 nm. At such a small $L_g$, quantum mechanical tunneling from the source to drain becomes appreciable. Several simulation works predicted that at $L_g \approx 5–10$ nm, $I_{SDT}$ will become prominent and severely degrade the subthreshold swing (SS) of MOSFETs [2,13-14]. Nonetheless, observation of SDT has been reported only in a few experiments, e.g. a Si MOSFET with $L_g = 8$ nm, using temperature-dependent measurements [28]. Whether the ultimate scaling limit of $L_g$ is set by $I_{SDT}$ is still not clear because of the lack of experimental evidence, and because the answer also depends on the precise geometry of the FET. However, to fully exploit the excellent electrostatic control of the ultra-thin CNTs, the $L_g$ of CNFETs is likely to be aggressively scaled down until the leakage current becomes intolerable. It is thus important to develop a model that takes into account the impact of $I_{SDT}$ in the sub-10-nm technology nodes.

Two tunneling mechanisms are considered here: SDT and band-to-band tunneling (BTBT) at the drain side. The SDT can be further divided into two parts: the intra-band SDT (intra-SDT), the tunneling from conduction band (CB) to CB, and the inter-band SDT (inter-SDT), the tunneling from CB to valence band (VB) to CB. The BTBT is the tunneling from source VB to drain CB, as illustrated in Fig. 5. While n-type FETs are used as examples throughout this paper, the model can be easily applied to p-FETs by properly changing the polarity of the terminal voltages, due to the symmetry of the CNT CB and VB. All tunneling currents are computed by the 1D Landauer formula [24]:

$$I = \frac{4q}{h} \int T_e(E) \left[ f(E, E_{fs}) - f(E, E_{fd}) \right] dE \quad (6)$$

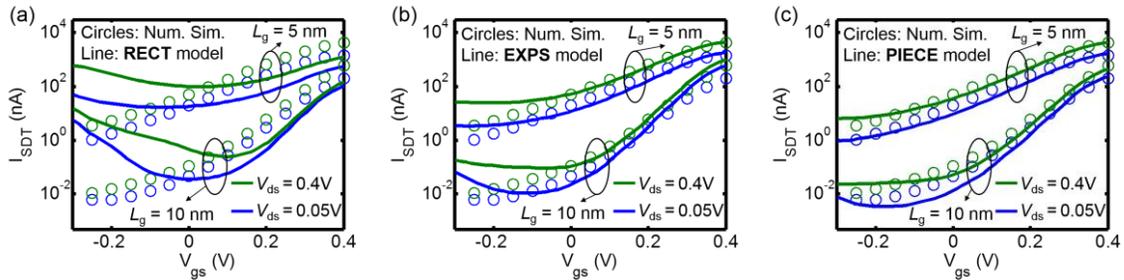

Fig. 7. Comparison of direct source-to-drain tunneling current between the numerical simulation [30] and the three models—(a) RECT: a rectangular $E_c$ profile; (b) EXPS: $E_c$ profile given by (9); and (c) PIECE: $E_c$ profile given by (10)—for different gate lengths. $d = 1$nm is used.

where $T_e$ is the tunneling probability, and $E_{fs}$ and $E_{fd}$ are Fermi levels at the source and the drain, respectively. $T_e$ is calculated by the Wentzel–Kramers–Brillouin (WKB) approximation [29]:

$$T_e(E) = \exp\left(-2\int_{x_i}^{x_o} \kappa \, dx\right) \quad (7)$$

$$\kappa = \frac{\pi E_g}{h\upsilon_F}\sqrt{1-\{1-2[E_c(x)-E]/E_g\}^2}$$

where $\kappa$ is the imaginary wave vector in CNTs, $\upsilon_F \approx 10^6$ m/s is the Fermi velocity, $x$ is the position along the CNFET channel, $x_i$ and $x_o$ are the positions where the electrons "tunnel" in and out the energy barrier, respectively (see Fig. 5). Eq. (7) is then recast as following for the convenience of calculations:

$$T_e(E) = \exp\left[-\frac{2\pi E_g}{h\upsilon_F}t_b(E)\right] \quad (8)$$

$$t_b(E) = \int_{x_i}^{x_o}\sqrt{1-\{1-2[E_c(x)-E]/E_g\}^2}\,dx$$

To calculate $T_e$, analytical models for $E_c(x)$ are first discussed.

The circles in Fig. 6 are the $E_c$ profile calculated by the numerical simulation based on the NEGF quantum transport [30], which simulates a CNFET with a cylindrical gate-all-around (GAA) device structure and heavily doped S/D extensions. Two features are observed in the simulated $E_c$ profile: (i) a curvy profile around the top of $E_c(x)$ and (ii) gradual tails extending into the S/D extensions. Three different analytical models of $E_c(x)$ are examined here: (i) a rectangular profile (named RECT in Fig. 6); (ii) two connected exponential functions to model the curvy top of $E_c(x)$ (named EXPS in Fig. 6):

$$E_c(x) = \begin{cases} E_{cs}(x) = u_s e^{-x/\lambda} + v_s, -L_g/2 - L_{of} < x < 0 \\ E_{cd}(x) = u_d e^{x/\lambda} + v_d, 0 < x < L_g/2 + L_{of} \end{cases} \quad (9)$$

where $u$'s and $v$'s are fitting coefficients, $\lambda$ is the electrostatic length scale discussed in [16], and $L_{of}$ is an empirical parameter functioning like an extension of the $L_g$ that captures the finite Debye length and the gate fringing field (see Fig. 6); (iii) a piecewise function to describe both the curvy top and the tails of $E_c(x)$ (named PIECE in Fig. 6):

$$E_c(x) = \begin{cases} E_{cs}(x) = b_s e^{(x+L_g/2)/\lambda_s} + c_s, x < -L_g/2 \\ E_{cg}(x) = a_1 e^{-x/\lambda} + a_2 e^{x/\lambda} + a_3, -L_g/2 < x < L_g/2 \\ E_{cd}(x) = b_d e^{-(x-L_g/2)/\lambda_d} + c_d, x > L_g/2 \end{cases} \quad (10)$$

where $a$'s, $b$'s, $c$'s, $\lambda_s$, and $\lambda_d$ are fitting coefficients. By substituting (9) and (10) into (8), $T_e$ can be calculated analytically. Derivation of the coefficients in (9) and (10) as well as the analytical expressions of $T_e$ in (8) are detailed in [31, Eq. (28)-(36)]. $I_{SDT}$ is then calculated by (6) numerically.

$I_{SDT}$ calculated by the numerical simulation [30] is compared against the three different $E_c(x)$ models individually in Fig. 7a-c. As shown in Fig. 7a, the RECT model does not fit the data well in the high $V_{gs}$ region (i.e. near-threshold), because it fails to capture the characteristic of the curvy top of $E_c$, resulting in an underestimate of $I_{SDT}$; in the low $V_{gs}$ region (i.e. deep subthreshold region), the RECT model overestimates $I_{SDT}$ due to the disregard of the tails of the $E_c$ profile; in Fig. 7b, the EXPS model fits the data well at high $V_{gs}$ but overestimates $I_{SDT}$ at low $V_{gs}$ because it also fails to capture the tails; finally in Fig. 7c, the PIECE model gives the best fitting result because it considers both the curvy top and the tails. However, the use of a piecewise function in (10) could potentially result in convergence issues when implemented in Verilog-A [32], because when a large-scale circuit is simulated in an environment like SPICE, extraordinarily large biases may be applied on the device terminals, which can potentially lead to discontinuities in (10). As a result, the EXPS model will be used to calculate $I_{SDT}$ in the following analysis. Although the EXPS model overestimates $I_{SDT}$ in the deep subthreshold region, it can still give accurate results in the subthreshold region and warn the user of an imminent significant impact of $I_{SDT}$ when the $L_g$ becomes too short. Besides, the EXPS model is more computationally efficient.

As shown in [33], the presence of $I_{SDT}$ significantly degrades the SS and increases the leakage power of CNFETs. To explore potential ways to lower $I_{SDT}$, Figs. 8a and 8b illustrate how $I_{SDT}$ is affected by $d$, $n_{sd}$, and the dielectric constant of the sidewall spacer $k_{spa}$ (see Fig. 1): (i) as shown in Fig. 8a, $I_{SDT}$ increases exponentially with $d$, because $\kappa$ in (7) is proportional to $E_g$. By utilizing small-diameter CNTs, tunneling leakage can be effectively mitigated, but it also leads to lower drive current due to larger $R_c$, and lower carrier mobility and velocity [16]; (ii) a decrease of $n_{sd}$ from 1 nm$^{-1}$ to 0.6 nm$^{-1}$ can reduce $I_{SDT}$ by a factor of 3.5, because as $n_{sd}$ decreases, the CB edge at the source is raised relative to the Fermi level, and thus less carriers are available to tunnel from the source through the barrier to the drain (see the Fig. 8a inset). However, lower $n_{sd}$ gives higher $R_{ext}$; (iii) as shown in the Fig. 8b inset, higher $k_{spa}$ results in stronger gate-to-extension fringe field and leads to a wider energy barrier. To model the effect of the fringe field caused by different $k_{spa}$'s, $L_{of}$ in (9) and implicitly in (10) is empirically related to $k_{spa}$ and the gate oxide thickness $t_{ox}$:

$$L_{of} = (0.0263 k_{spa} + 0.056) \cdot t_{ox} \quad (11)$$

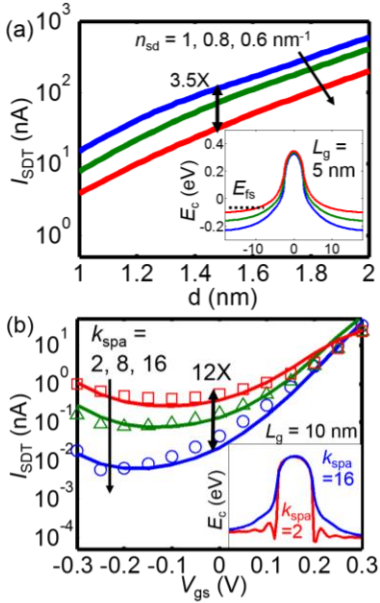

Fig. 8. (a) Direct source-to-drain tunneling current $I_{SDT}$ vs. CNT diameters for different doping density in S/D extensions. Inset: the source CB is raised as $n_{sd}$ decreases. (b) $I_{SDT}$ vs. $V_{gs}$ for different spacer dielectric constants ($k_{spa}$). Symbol: numerical simulation; Line: model. Inset: higher $k_{spa}$ results in stronger gate-to-extension fringe field, wider energy barrier, and lower $I_{SDT}$.

As shown in Fig. 8b, increasing $k_{spa}$ from 2 to 16 can reduce $I_{SDT}$ by a factor of 12 for $L_g = 10$ nm and $d = 1$ nm. However, increasing $k_{spa}$ also causes larger parasitic capacitances and degrades the circuit speed [34]. These results indicate that lowering $I_{SDT}$ may degrade the speed performance (i.e. increase delay), a manifestation of the energy-delay trade-offs. Note that (11) is a first-order approximation and the empirical coefficients are determined by fitting the $I_{SDT}$ model to the numerical simulation based on a GAA cylindrical structure [30] for different $k_{spa}$ and $t_{ox}$. While (11) could be changed for different device geometries, the trend should remain the same.

The BTBT current ($I_{BTBT}$) is modeled in a similar approach to $I_{SDT}$, except that the $E_c$ is modeled differently:

$$E_c(x) = u e^{-x/\lambda_{BTBT}} \quad (12)$$

where $u$ and $\lambda_{BTBT}$ are fitting parameters. Eq. (12) is employed to model the decaying $E_c$ profile at the gate-drain junction (see Fig. 5). Substituting (12) into (8) gives:

$$t_b(E) = \int_{x_i}^{x_o} \sqrt{1 - \{1 - 2(u e^{-x/\lambda_{BTBT}} - E)/E_g\}^2}\, dx \quad (13)$$

$$x_i = \lambda_{BTBT} \ln\left(\frac{E+E_g}{u}\right), x_o = \lambda_{BTBT} \ln\left(\frac{E}{u}\right)$$

By changing variables, a closed-form expression of $t_b$ is obtained:

$$t_b = \lambda_{BTBT} \pi \left(\zeta + \sqrt{\zeta^2 - 1}\right) \quad (14)$$

where $\zeta = -2E/E_g - 1$ (see [31] for detailed derivation). $I_{BTBT}$ is then obtained by integrating (6) numerically. The modeled $I_{BTBT}$ is compared against the numerical simulation in Fig. 9a and 9b. Similar to the discussion of the effect of gate-to-drain fringe fields when modeling $I_{SDT}$, $I_{BTBT}$ is also a function of $k_{spa}$: The higher the $k_{spa}$, the stronger the fringe fields, the more gradual the $E_c$ profile at the gate-drain junction, and the smaller the $I_{BTBT}$. Empirically, $\lambda_{BTBT}$ (nm) = $0.092 k_{spa} + 2.13$ is determined by fitting the $I_{BTBT}$ model to the numerical simulation result. Note that phonon-assisted and trap-assisted tunneling [35] are not considered in this model, so $I_{BTBT} = 0$ when $V_{ds} < E_g$. In addition, since the tunneling model presented in this paper are calibrated to the NEGF-based numerical simulation with a relatively simple GAA cylindrical device structure [30] assuming ballistic transport, the model aims to provide a trend instead of accurate results.

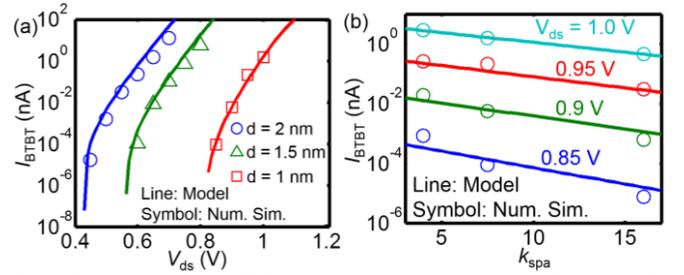

Fig. 9. Calibration of the BTBT current model to the numerical simulation [30] for different CNT diameters and spacer dielectric constants $k_{spa}$. (a) $I_{BTBT}$ vs $V_{ds}$ for different diameters. (b) $I_{BTBT}$ vs $k_{spa}$ for different $V_{ds}$'s.

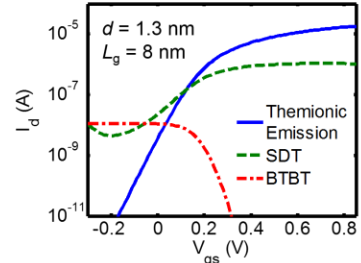

Fig. 10. A representative $I_d$ vs. $V_{gs}$ of a CNFET with $L_g = 8$ nm and $d = 1.3$ nm, showing that the tunneling currents dominate over the thermionic emission current in the subthreshold region.

## IV. CNFET PERFORMANCE ASSESSMENT

The intrinsic elements of the VS-CNFET model introduced in [16] is then combined with the extrinsic elements described in Sections II and III to assess the CNFET design space and performance. A representative $I_d$ vs. $V_{gs}$ curve given by the complete VS-CNFET model separately identifying the current components—thermionic emission, direct SDT, and BTBT currents—is shown in Fig. 10. It can be seen that tunneling currents can dominate over the thermionic emission current in the subthreshold region of a short-channel CNFET.

In this section, we demonstrate the capability of the VS-CNFET model by optimizing $L_g$, $L_c$, $L_{ext}$, and CNT diameter to minimize the CNFET gate delay ($\tau_{gate}$) and estimating the requirement for CNT density ($\rho_{cnt} \equiv 1/s$, where $s$ is the spacing between CNTs. see Fig. 1) to meet the ITRS targets of drive current. For advanced CMOS technology, the dimensional scaling is no longer simply the scaling of $L_g$ but a multi-variable optimization that targets a technology pacing objective. Fig. 11 shows the dimensional scaling trend of major foundries as well as the projections down to the so-called 5-nm technology node by linear extrapolation. While foundries tend to scale the metal-

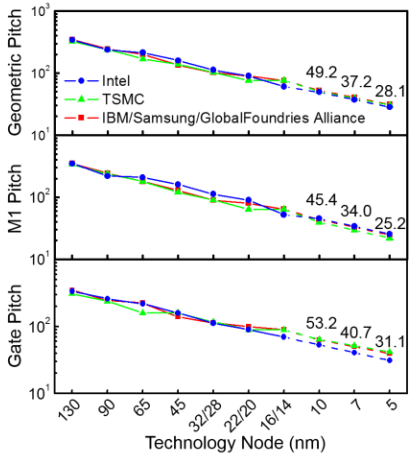

Fig. 11. Dimensional scaling trend of major foundries collected from the published data (unit in nm). The "geometric" pitch is defined as (metal-1 pitch × contacted gate pitch)$^{1/2}$. The dashed lines beyond the 16/14-nm node are projections by linearly extrapolation from the nodes over the last 10 years.

1 pitch ($L_{M1}$) and the contacted gate pitch ($L_{pitch}$, illustrated in Fig. 1) at different paces, the "geometric" pitch $L_{GP} \equiv (L_{M1} \cdot L_{pitch})^{1/2}$ scales at a relatively consistent pace. Here we use this $L_{GP}$ to pace the advancement of logic technology. The CNFET performance is evaluated at the 5-nm node corresponding to $L_{GP} = 28.1$ nm, $L_{M1} = 25.2$ nm, and $L_{pitch} = 31.1$ nm; the "2023" node of the 2013 ITRS projections [17] is used as a reference point, which also predicts $L_{M1}$ will be scaled down to 25.2 nm in 2023 for high performance logic. The corresponding ITRS parameters—supply voltage $V_{dd} = 0.71$ V, and EOT = 0.51 nm—are used as the inputs to the VS-CNFET model. Furthermore, a GAA device structure is assumed (see Fig. 1) in the following analysis.

Under the constraint of a fixed $L_{pitch}$, trade-offs exists between $L_g$, $L_c$, and $L_{ext}$ at the device-level: (i) scaling down $L_g$ helps to improve the device speed because of lower intrinsic capacitance and higher drive current, but also increases the off-state current ($I_{off}$, defined as the $I_d$ at $V_{gs} = 0$ and $V_{ds} = V_{dd}$) and thus the static power. Hence there exists an optimal $L_g$ to balance the speed and power consumption; (ii) $L_c$ is preferred to be as long as possible in order to lower the $R_c$ (ignoring the possible increase in the parasitic capacitance at the circuit-level); (iii) scaling down $L_{ext}$ helps to reduce $R_{ext}$ but drastically increase the parasitic capacitance ($C_{par}$). For CNFETs, $R_{ext}$ is negligible compared to $R_c$ in general, so $L_{ext}$ is preferred to be large.

In Fig. 12, $L_g$, $L_c$, and $L_{ext}$ are optimized under the constraints of $L_{pitch} = 31$ nm and $I_{off} = 100$ nA/μm (by adjusting the flat-band voltage $V_{fb}$) to minimize $\tau_{gate} \equiv (L_g C_{inv} + C_{par}) \cdot V_{dd}/I_{on}$, where $C_{par}$ is calculated by the analytical models of [36], in which the gate-to-extension fringe capacitance ($C_{of}$) and gate-to-contact capacitances ($C_{gtc}$) are considered (see Fig. 1). $\rho_{cnt} = 100$ CNTs/μm is assumed. The optimal design is arrived at $L_g = 11.7$ nm, $L_c = 12.9$ nm, and $L_{ext} = 3.2$ nm. Because the optimization goal is to minimize $\tau_{gate}$ and $R_c$ is the major limiter of the drive current, $L_g$ is scaled down until $I_{off}$ becomes intolerable and $L_{ext}$ is scaled down until $C_{par}$ becomes too large, in order to save space for $L_c$. It is worthwhile noting that while the optimal design may vary as different parameters (e.g. CNT diameter) are used, the shape of the contour in Fig. 12 remains the same.

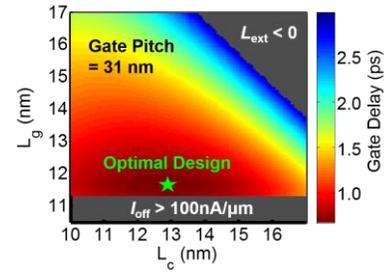

Fig. 12. Optimization of the CNFET dimensions ($L_g$, $L_c$, and $L_{ext}$) to minimize the gate delay under the constraints of $L_{pitch} = 31$ nm and $I_{off} = 100$ nA/μm. $\rho_{cnt} = 100$ CNTs/μm and $d = 1.2$ nm are used.

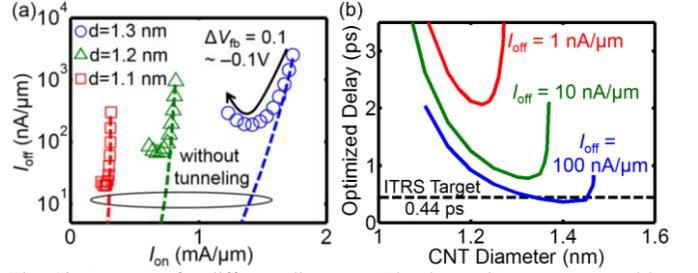

Fig. 13. $I_{on}$ vs. $I_{off}$ for different diameters. The data points are generated by changing $\Delta V_{fb}$ from –0.1 V to 0.1 V; the dashed lines represents the case when the tunneling current is turned off; the smaller $I_{on}$ for small d is mainly due to the larger $R_c$. $\rho_{cnt} = 100$ CNTs/μm is assumed.

It appears in Fig. 12 here that $L_g$ cannot scale below 11 nm in order to keep $I_{off} > 100$ nA/μm, mainly due to SDT. Since SDT highly depends on CNT diameter, the impact of CNT diameter is studied in Fig. 13: Fig. 13a shows $I_{on}$ vs. $I_{off}$ for different diameters. A minimum $I_{off}$ for each $d$ is observed by sweeping $V_{fb}$: as $V_{fb}$ starts increasing, $I_{off}$ decreases exponentially because both thermionic emission and intra-SDT currents decrease; as $V_{fb}$ further increases beyond a certain point, inter-SDT starts to increase and becomes dominant, so $I_{off}$ increases. The larger the diameter, the higher the $I_{off}$. In addition, for small-diameter CNTs, reducing $V_{fb}$ does not improve $I_{on}$ effectively, because the $R_c$ is so large that the $I_{on}$ is dominated by the resistance of contacts rather than the channel. In Fig. 13b, we co-optimize CNT diameter, $L_g$, $L_c$, and $L_{ext}$ to minimize $\tau_{gate}$ under different constraints of $I_{off}$. Each point along the curves has different optimized $L_g$, $L_c$, and $L_{ext}$. The optimal diameter increases as the constraint of $I_{off}$ increases, indicating that large-diameter CNTs are suitable for high performance applications while small-diameter CNTs are suitable for low power applications.

In the discussion above, the CNTs are assumed to be perfectly aligned and equally spaced, and $\rho_{cnt} = 100$ CNTs/μm is assumed. This CNT density is within reach experimentally as suggested in recent reports: the highest $\rho_{cnt}$ to date through chemical vapor deposition (CVD) is ≈ 30 CNTs/μm [37]; by using multiple CNT transfers, $\rho_{cnt} \approx 100$ CNTs/μm was achieved [38]; although $\rho_{cnt} > 500$ CNTs/μm has been reported in [39] by assembling solution-based CNTs using the Langmuir-Schaefer method on a target substrate, the CNTs were not well-aligned and the measured $R_c \approx 3$ MΩ/CNT, about 100× the value reported in [23]. While high $\rho_{cnt}$ has been reported in these works, the control of CNT pitch still remains to be a challenge. Variations in CNT pitch can degrade CNFET performance and reduce circuit yield. The issue of CNT variations has been discussed in [10] and is out of the scope of this paper.

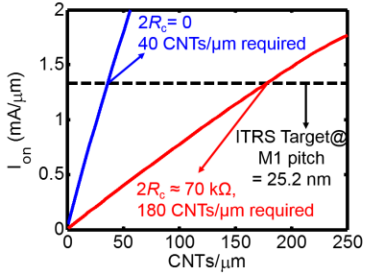

Fig. 14. Projection of the requirement for CNT density to meet the 2013 ITRS target of $I_{on}$ = 1.33 mA/μm with fixed $I_{off}$ = 100 nA/μm corresponding to metal-1 pitch = 25.2 nm. $2R_c \approx$ 70 kΩ per CNT is calculated by (2) with $L_c$ = 12.9 nm and $d$ = 1.2 nm.

To estimate the $\rho_{cnt}$ required for CNFETs to deliver enough drive current (assuming no variations), Fig. 14 shows $I_{on}$ vs. $\rho_{cnt}$ with a fixed $I_{off}$ = 100 nA/μm; $d$ = 1.2 nm is used for the analysis because it is the diameter measured in the experiments that the model is calibrated to [16]; $L_g$ = 11.7 nm and $L_c$ = 12.9 nm are used according to the optimization result from Fig. 12. At $L_c$ = 12.9 nm, $2R_c \approx$ 70 kΩ per CNT, and $\rho_{cnt} \approx$ 180 CNTs/μm is needed in order to meet the 2013 ITRS target of $I_{on}$ = 1.33 mA/μm (corresponding to $L_{M1}$ = 25.2 nm); whereas when $R_c$ can be reduced to zero, the required $\rho_{cnt}$ can be lowered to 40 CNTs/μm.

## V. DISCUSSION

The analysis in Section IV exhibits the potential of scalability of CNFETs down to $L_{pitch}$ = 31 nm and capability of delivering high drive current with on/off ratio > $10^4$. It is important to review the assumptions made in the analysis: (i) the interface between the gate dielectric and the CNTs are assumed to be perfect, i.e. hysteresis of the I-V characteristics [8] is negligible, and the short-channel effect (e.g. SS degradation and DIBL) is determined purely by electrostatics. Recent progress in CNT-dielectric interface includes the use of $Y_2O_3$ and $LaO_3$ as gate dielectrics to reduce the interface traps [40-41] and interface passivation to alleviate the hysteresis [8,42]. (ii) The CNTs are assumed to be perfectly aligned and equally spaced. The imperfect alignment and variation in the CNT spacing result in delay variations and potential functional failures. Process techniques to achieve good CNT alignment have been improved over the years [43]; design techniques can be employed to overcome these imperfections at modest cost of area and energy consumption [10]. Nonetheless, improvement in the material is still strongly desired. (iii) The CNTs in a single device are assumed to be identical in diameter, carrier mobility and velocity. However, Cao *et al.* measured the distribution of CNT diameter and mobility [9], showing that the variations are not negligible. As these imperfections are considered, the projections described in Section IV need to be adjusted, but the general conclusion should remain unchanged (e.g. trade-off between contact resistance and tunneling currents due to the selection of CNT diameter).

Since CNT diameter is shown to have great impact on $R_c$, $I_{SDT}$, and thus the CNFET performance, we next revisit the model and discuss its validity. The dependence of $R_c$ on $d$ is characterized in (3) by $E_{00}$, which can be viewed (loosely) as the inverse of the Schottky barrier width at the metal-CNT contacts. Smaller $E_{00}$ leads to higher sensitivity of $R_c$ to CNT diameter. In this paper, $E_{00}$ = 32 meV is extracted from [20]. However, detailed experimental studies on the dependence of $R_c$ on $d$ are still lacking, and whether small-diameter CNTs will lead to such a large $R_c$ (see Fig. 3) that the drive current of CNFETs becomes too small for practical applications needs to be verified by more careful investigation. On the other hand, though large-diameter CNTs can give lower $R_c$, it also causes high tunneling leakage current. As shown in Fig. 8, $I_{SDT}$ increases drastically as $d$ increases. The model of tunneling currents developed in Section III is calibrated to the numerical simulation [30]. However, to date, only a few experimental works have observed $I_{SDT}$ in Si-MOSFET with $L_g$ = 8 nm [28], and experimental observation of $I_{SDT}$ in CNFETs has not been reported yet. For a CNFET with $L_g$ = 9 nm and $d \approx$ 1.3 nm, as reported in [3], $I_{SDT}$ is expected to be appreciable, but has not yet been clearly observed. One manifestation of $I_{SDT}$ is the degradation of inverse subthreshold slope (SS). Temperature-dependent measurement of SS can be helpful to identify the existence of $I_{SDT}$: if $I_{SDT}$ is not prominent, the SS will decrease as the temperature goes down; if $I_{SDT}$ is significant, the SS will not decrease but remain relatively unchanged as the temperature goes down, as described in [28]. Since large-diameter CNTs can provide higher drive current, research on whether the tunneling current in scaled CNFETs is tolerable or not is of crucial importance, and temperature-dependent measurement is suggested to be an effective means to identify the existence of $I_{SDT}$.

## VI. CONCLUSION

We present data-calibrated analytical models for metal-CNT contact resistance, direct source-to-drain and band-to-band tunneling leakage currents in CNFETs, which are integrated with the intrinsic model elements to arrive at a complete CNFET model for performance assessment. We predict that a density of 180 CNTs/μm is required to meet the ITRS targets of off-state and on-state currents at the 5-nm technology node corresponding to 25.2 nm metal-1 pitch and 31 nm contacted gate pitch assuming no variations; in contrast, a density of 40 CNTs/μm would be enough if the parasitic contact resistance can be eliminated. Experimental demonstrations of >100 CNTs/μm are available today [38], but whether these are sufficient for highly scaled CNFETs remains to be seen, depending on $R_c$ optimization and diameter selection, as discussed in this study. In-depth study of $R_c$ and its dependence on $d$ is highly desirable in order to identify further device design points for CNFET technology in the sub-10-nm nodes.

## ACKNOWLEDGMENT

The authors would like to thank Prof. L. Wei (Waterloo), Prof. S. Rakheja (NYU), G. Hills (Stanford), Prof. S. Mitra

(Stanford), and Prof. Z. Chen (Purdue) for the useful discussion. This work was supported in part through the NCN-NEEDS program, which is funded by the National Science Foundation, contract 1227020-EEC, and by the Semiconductor Research Corporation, and through Systems on Nanoscale Information fabriCs (SONIC), one of the six SRC STARnet Centers, sponsored by MARCO and DARPA, the member companies of the Initiative for Nanoscale Materials and Processes (INMP) at Stanford University, as well as IBM through the Center for Integrated Systems (CIS) at Stanford University.